\documentclass[11pt]{article}

\title{Tiling 3D Space by (Translates of) a Single Polycube is Undecidable}%
\author{%
  Erik D. Demaine%
\footnote{\protect\label{footnote:MIT-CSAIL}%
    Computer Science and Artificial Intelligence Laboratory,
    Massachusetts Institute of Technology, Cambridge, MA, USA,
    \protect\href{mailto:edemaine@mit.edu}{\protect\nolinkurl{edemaine@mit.edu}}}
  \and
  Stefan Langerman%
\footnote{\protect\label{footnote:ULB}%
    Computer Science Department, Universit\'e libre de Bruxelles, Belgium,
    \protect\href{mailto:stefan.langerman@ulb.ac.be}{\protect\nolinkurl{stefan.langerman@ulb.ac.be}}. S. Langerman is Directeur de Recherche du F.R.S.-FNRS.}
}%

\date{}

  \usepackage[T1]{fontenc}
  \usepackage{libertine}
  \usepackage[margin=1in]{geometry}
  \usepackage{amsmath}
  \usepackage{amssymb}
  \usepackage{amsthm}
  \usepackage{mathtools}
  \usepackage[libertine]{newtxmath}
  \usepackage[scaled=0.96]{zi4}
  \usepackage{upquote}
  \usepackage{mathrsfs}
  \usepackage{microtype}
  \usepackage{enumitem}
  \usepackage{booktabs}
  \usepackage{tabularx}
  \usepackage{graphicx}
  \usepackage{subcaption}
  \usepackage[table]{xcolor}
  \usepackage{tikz}
  \usepackage{listings}
  \usepackage{tcolorbox}
  \usepackage{aliascnt}
  \usepackage{xurl}
  \usepackage[numbers,sort&compress]{natbib}
  \usepackage{hyperref}
  \usepackage[capitalise,nameinlink,noabbrev]{cleveref}
\usetikzlibrary{arrows.meta,positioning,calc}
\lstdefinelanguage{lean}{
  morekeywords={import,abbrev,def,theorem,fun,by},
  morekeywords=[2]{Prop,Type,Set,List,Fin,Nat,Bool},
  sensitive=true, morecomment=[l]{--}, morestring=[b]''}
\lstdefinestyle{lean}{language=lean, basicstyle=\ttfamily\small, columns=fullflexible,
  keepspaces=true, showstringspaces=false, extendedchars=false,
  breaklines=true, breakatwhitespace=true,
  keywordstyle=\bfseries\color{figblue}, keywordstyle=[2]\color{figgreen},
  literate={ℕ}{{$\mathbb{N}$}}1 {ℤ}{{$\mathbb{Z}$}}1 {ℝ}{{$\mathbb{R}$}}1 {×}{{$\times$}}1 {→}{{$\to$}}1
    {∀}{{$\forall$}}1 {∃}{{$\exists$}}1 {∈}{{$\in$}}1
    {⊆}{{$\subseteq$}}1 {∧}{{$\wedge$}}1 {∨}{{$\vee$}}1 {≤}{{$\le$}}1 {≠}{{$\ne$}}1 {↔}{{$\leftrightarrow$}}1
    {≃ᵃⁱ}{{$\simeq^{\mathrm{ai}}$}}3
    {¬}{{$\neg$}}1 {∂}{{$\partial$}}1 {∑}{{$\sum$}}1 {ᵐ}{{$^{\mathrm m}$}}1}
\tcbuselibrary{skins}
\newtcolorbox{leanbox}{colback=black!3,colframe=black!45,boxrule=0.5pt,arc=2pt,
  left=6pt,right=6pt,top=3pt,bottom=3pt,enhanced,
  overlay={\node[anchor=north east,inner sep=3pt,draw=black!45,fill=white,line width=0.5pt,
    rounded corners=2pt] at (frame.north east)
    {\includegraphics[height=1.6ex]{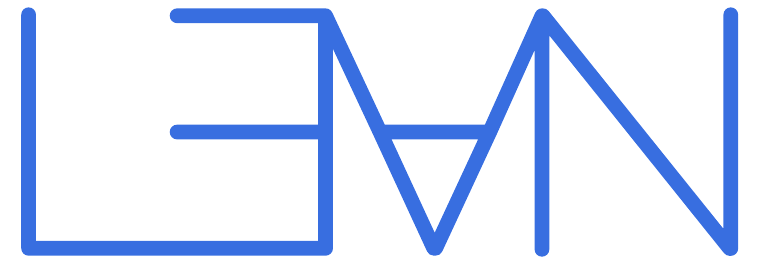}};}}
\newcommand{\leanlogo}{\raisebox{-0.15ex}{\includegraphics[height=1.5ex]{figures/lean-logo.pdf}}}
\newcommand{\leanitem}[1]{\leansep\lstinputlisting[style=lean]{lean-snippets/#1.lean}\def\leansep{\vspace{2pt}}}
\newcommand{\leansnippet}[1]{\begin{leanbox}\def\leansep{}\let\do\leanitem\docsvlist{#1}\end{leanbox}}
\definecolor{figblue}{RGB}{41,110,166}
\definecolor{figorange}{RGB}{217,107,38}
\definecolor{figgreen}{RGB}{38,128,89}
\definecolor{bgblue}{HTML}{C1C1F6}
\definecolor{bgred}{HTML}{F6C1C1}
\definecolor{bgyellow}{HTML}{FFF5C4}
\definecolor{bggreen}{HTML}{CDEECF}
\newtheorem{theorem}{Theorem}[section]
\newaliascnt{lemma}{theorem}

\aliascntresetthe{lemma}
\newaliascnt{proposition}{theorem}

\aliascntresetthe{proposition}
\newaliascnt{corollary}{theorem}

\aliascntresetthe{corollary}
\theoremstyle{definition}
\newaliascnt{definition}{theorem}

\aliascntresetthe{definition}
\theoremstyle{remark}
\newaliascnt{remark}{theorem}

\aliascntresetthe{remark}
\crefname{lemma}{Lemma}{Lemmas}
\crefname{proposition}{Proposition}{Propositions}
\crefname{corollary}{Corollary}{Corollaries}
\crefname{definition}{Definition}{Definitions}
\crefname{remark}{Remark}{Remarks}
\newcommand{\Z}{\mathbb Z}

\newcommand{\R}{\mathbb R}

\newcommand{\calT}{\mathcal T}

\numberwithin{equation}{section}
\setlist{itemsep=3pt,topsep=5pt}
\let\realbfseries=\bfseries
\def\bfseries{\realbfseries\boldmath}

\providecommand\defn[1]{\textbf{\textit{\boldmath #1}}}

\begin{document}
\maketitle
\begin{abstract}
  We prove co-RE-completeness, and thus undecidability, of the following
  problem: given a single (connected) polycube,
  decide whether it tiles 3D Euclidean space by translations.
  The same result holds if we also allow rotations and/or reflections.
  We reduce from Wang tiling using the decorated two-prime Sudoku construction of Greenfeld and Tao and a cyclic encoding adapted from OpenAI's 3D aperiodic tile, and apply a reduction of Kim to make the prototile connected (via faces).
  Dimension three is optimal: translational monotiling is known to be
  decidable in $\Z^2$ and for a single (possibly disconnected) polyomino in $\R^2$.
\end{abstract}

\begin{tcolorbox}[colback=figblue!6!white,colframe=figblue!65!black,boxrule=0.6pt,arc=2pt,left=10pt,right=10pt,top=8pt,bottom=8pt]
  This paper gives a human-readable overview of the
  results, proof ideas, and formal theorem statements. The proof is verified
  in Lean. See \cref{sec:approach} for details of our approach, and
  \protect\href{https://github.com/edemaine/translation-tiling-paper}{the repository}
  \cite{TilingRepo}
  for the complete Lean code and a longer LLM-written technical article.
\end{tcolorbox}

\section{Introduction \label{sec:intro}}

\begin{figure}[htbp]
  \centering
  \begin{tikzpicture}[>=Latex]
    \node[inner sep=0pt,outer sep=0pt] (prototile)
      {\includegraphics[height=3.3cm]{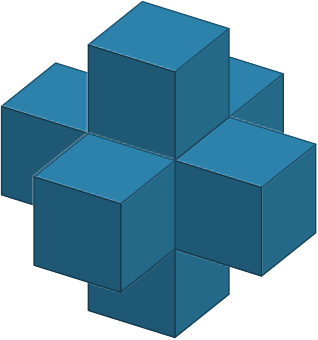}};
    \node[inner sep=0pt,outer sep=0pt,right=2.1cm of prototile] (tiling)
      {\includegraphics[height=5.2cm]{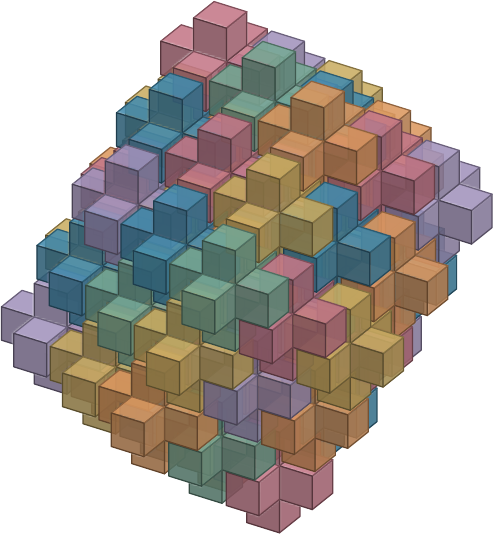}};
    \draw[->,semithick]
      ([xshift=3mm]prototile.east) -- node[above] {$?$} ([xshift=-3mm]tiling.west);
  \end{tikzpicture}
  \caption{%
    The tiling-by-translation decision problem:
    given $k$ shapes, can translated copies tile $d$D space?
    Here, $k=1$ and $d=3$, the main version solved in this paper.}%
  \label{fig:tiling-decision}
\end{figure}

Given $k$ finite subsets of $\Z^d$ (\defn{prototiles}),
can (infinitely many) copies of those subsets be \emph{translated}
to cover all of $\Z^d$ without overlaps?
(Refer to \cref{fig:tiling-decision}.)
For which values of $k$ and $d$ is this problem decidable?
In this paper, we prove undecidability for $k=1$ and $d=3$.

The decidability question has a long history, summarized in \cref{tab:related}.
The first result is Golomb's implementation \cite{Golomb}
of the undecidable Wang tiling problem \cite{Berger},
which works in 2D but requires arbitrarily many prototiles.
In 2D, Ollinger \cite{Ollinger} reduced the number of prototiles to a constant
(11), which was subsequently improved in a series of papers;
the current record is 3 by the MIT--ULB CompGeom Group \cite{Trominoes}.
In 3D, the first undecidability result used 6 prototiles
\cite{YangSix},
and the current record is 2 \cite{Kim};
we improve to the minimum possible number of 3D prototiles, 1.
3D is also the minimum dimension for undecidability with 1 prototile,
given known algorithms for 2D \cite{BN91,Bhattacharya,GTStructure}.

\begin{table}[tp]
  \centering
  \small
  \renewcommand{\arraystretch}{1.3}
  \begin{tabularx}{\textwidth}{c
    >{\hsize=0.75\hsize\linewidth=\hsize\centering\arraybackslash}X
    >{\hsize=1.25\hsize\linewidth=\hsize\centering\arraybackslash}X
    *{2}{>{\centering\arraybackslash}X}}
    \toprule
    \multicolumn{5}{c}{\textbf{Translations only}} \\
    \midrule
    Prototiles & 1D & 2D & 3D & $d$D ($d\geq4$) \\
    \midrule
    \rowcolor{bggreen}\cellcolor{white}%
    1 & R ($\downarrow$) & R \cite{BN91,Bhattacharya,GTStructure} & \cellcolor{bgblue}co-RE-c (Thm.~\ref{thm:undecidability})$^{*}$ & \cellcolor{bgblue}co-RE-c ($\leftarrow$)$^{*}$ \\
    \rowcolor{bgred}\cellcolor{white}%
    2 & \cellcolor{bggreen}R ($\downarrow$) & \cellcolor{bgyellow}OPEN & co-RE-c \cite{Kim} ($\uparrow$)$^{*}$ & co-RE-c ($\uparrow$)$^{*}$ \\
    \rowcolor{bgred}\cellcolor{white}%
    3 & \cellcolor{bggreen}R ($\downarrow$) & co-RE-c \cite{Trominoes} & co-RE-c \cite{YangThree,Trominoes} ($\uparrow$)$^{*}$ & co-RE-c ($\uparrow$)$^{*}$ \\
    \rowcolor{bgred}\cellcolor{white}%
    4 & \cellcolor{bggreen}R ($\downarrow$) & co-RE-c \cite{YZFour} ($\uparrow$) & co-RE-c ($\uparrow$)$^{*}$ & co-RE-c ($\uparrow$)$^{*}$ \\
    \rowcolor{bgred}\cellcolor{white}%
    5 & \cellcolor{bggreen}R ($\downarrow$) & co-RE-c \cite{KimPlane}$^{*}$ & co-RE-c ($\uparrow$)$^{*}$ & co-RE-c ($\uparrow$)$^{*}$ \\
    \rowcolor{bgred}\cellcolor{white}%
    6 & \cellcolor{bggreen}R ($\downarrow$) & co-RE-c ($\uparrow$)$^{*}$ & co-RE-c \cite{YangSix} ($\uparrow$)$^{*}$ & co-RE-c ($\uparrow$)$^{*}$ \\
    \rowcolor{bgred}\cellcolor{white}%
    7 & \cellcolor{bggreen}R ($\downarrow$) & co-RE-c \cite{YZSeven,YZConvex} ($\uparrow$)$^{*}$ & co-RE-c ($\uparrow$)$^{*}$ & co-RE-c ($\uparrow$)$^{*}$ \\
    \rowcolor{bgred}\cellcolor{white}%
    8 & \cellcolor{bggreen}R ($\downarrow$) & co-RE-c \cite{YZEight} ($\uparrow$)$^{*}$ & co-RE-c ($\uparrow$)$^{*}$ & co-RE-c ($\uparrow$)$^{*}$ \\
    \rowcolor{bgred}\cellcolor{white}%
    9 & \cellcolor{bggreen}R ($\downarrow$) & co-RE-c \cite{YangNine} ($\uparrow$)$^{*}$ & co-RE-c ($\uparrow$)$^{*}$ & co-RE-c ($\uparrow$)$^{*}$ \\
    \rowcolor{bgred}\cellcolor{white}%
    10 & \cellcolor{bggreen}R ($\downarrow$) & co-RE-c \cite{YangTen} ($\uparrow$)$^{*}$ & co-RE-c ($\uparrow$)$^{*}$ & co-RE-c ($\uparrow$)$^{*}$ \\
    \rowcolor{bgred}\cellcolor{white}%
    11 & \cellcolor{bggreen}R ($\downarrow$) & co-RE-c \cite{Ollinger} ($\uparrow$)$^{*}$ & co-RE-c ($\uparrow$)$^{*}$ & co-RE-c ($\uparrow$)$^{*}$ \\
    \rowcolor{bgred}\cellcolor{white}%
    $n$ & \cellcolor{bggreen}R \cite{GTUndecidable} & co-RE-c \cite{Berger,Golomb} ($\uparrow$)$^{*}$ & co-RE-c ($\uparrow$)$^{*}$ & co-RE-c ($\uparrow$)$^{*}$ \\
    \bottomrule
  \end{tabularx}

  \caption{Complexity of tiling the whole integer lattice by translations with
    possibly disconnected polycubes, organized as in \cite[Table~3]{Trominoes}.
    Rows bound the number of prototiles; $n$ means an arbitrary finite
    input set. The last column applies to every fixed $d\geq4$.
    ``R'' (green) denotes decidability,
    ``co-RE-c'' (red) denotes co-RE-completeness,
    and yellow marks open cases.
    Blue highlights our new one-tile result
    (\cref{thm:undecidability}) and its extension to higher dimensions by extrusion.
    Results marked ``$^*$'' work even when every prototile is connected.
    (Unmarked results allow disconnected tiles.)
    Arrows point to adjacent entries that imply the stated result,
    including the connectedness restriction when marked.
    }
  \label{tab:related}
\end{table}

\begin{theorem}\label{thm:undecidability}
  For every fixed $d\geq3$, tiling $\Z^d$ by translations of a single finite
  connected%
\footnote{\protect\label{footnote:connected}%
    Connectivity of a set in $\Z^d$ is defined according
    to lattice adjacency in $\Z^d$.
    Connectivity of a polycube in $\R^d$ is thus defined according
    to facet adjacency of the constituent unit cubes.
    See \cref{sec:lean-connected}.}%
 subset of $\Z^d$ is co-RE-complete,%
\footnote{\protect\label{footnote:lean-complete}%
    \leanlogo{} theorem \texttt{tiling\char`\_coRE\char`\_complete}. See \cref{sec:lean-z3}.}
  and hence undecidable.%
\footnote{\protect\label{footnote:lean-undecidable}%
    \leanlogo{} theorem \texttt{tiling\char`\_undecidable}. See \cref{sec:lean-z3}.}
  Tiling $\R^d$ with a single finite connected\footref{footnote:connected} polycube,
  using any set of Euclidean motions containing all translations,
  is co-RE-hard,%
\footnote{\protect\label{footnote:lean-rigid}%
    \leanlogo{} theorem \texttt{real\char`\_tiling\char`\_coRE\char`\_hard}. See \cref{sec:lean-rigid}.}%
 and hence undecidable.%
\footnote{\protect\label{footnote:lean-real-undecidable}%
    \leanlogo{} theorem \texttt{real\char`\_tiling\char`\_undecidable}. See \cref{sec:lean-rigid}.}
  With translations and optionally rotations and/or reflections,
  the problem is co-RE-complete.%
\footnote{\protect\label{footnote:lean-real}%
    \leanlogo{} theorem \texttt{real\char`\_tiling\char`\_coRE\char`\_complete}. See \cref{sec:lean-r3}.}
\end{theorem}

\defn{Co-RE} is the complexity class of decision problems whose ``no'' instances
can be recognized in finite time by an algorithm (Turing machine).
A problem is \defn{co-RE-hard} if every co-RE problem can be reduced to it
via a computable many-one reduction,
and \defn{co-RE-complete} if it is also in co-RE.
It is known that some co-RE-complete problems
(such as ``does a given algorithm run forever?'')\ are undecidable,
so co-RE-completeness implies undecidability.

Our results also improve the state-of-the-art for tiling with translations
and rotations (and optionally reflections), as summarized in \cref{tab:rotations}.
Previously, the best undecidability result for 3D required two prototiles
\cite{Trominoes}.
The 2D case with one prototile remains open.

\begin{table}[tp]
  \centering
  \small
  \renewcommand{\arraystretch}{1.3}
  \begin{tabularx}{\textwidth}{c
    >{\hsize=0.75\hsize\linewidth=\hsize\centering\arraybackslash}X
    >{\hsize=1.25\hsize\linewidth=\hsize\centering\arraybackslash}X
    *{2}{>{\centering\arraybackslash}X}}
    \toprule
    \multicolumn{5}{c}{\textbf{Translations + rotations (+ reflections)}} \\
    \midrule
    Prototiles & 1D & 2D & 3D & $d$D ($d\geq4$) \\
    \midrule
    \rowcolor{bggreen}\cellcolor{white}%
    1 & R ($\downarrow$) & \cellcolor{bgyellow}OPEN & \cellcolor{bgblue}co-RE-c (Thm.~\ref{thm:undecidability})$^*$ & \cellcolor{bgblue}co-RE-c (Thm.~\ref{thm:undecidability})$^*$ \\
    \rowcolor{bgred}\cellcolor{white}%
    2 & \cellcolor{bggreen}R ($\downarrow$) & co-RE-c \cite{Trominoes} & co-RE-c \cite{Trominoes} ($\uparrow$)$^*$ & co-RE-c ($\uparrow$)$^*$ \\
    \rowcolor{bgred}\cellcolor{white}%
    3 & \cellcolor{bggreen}R ($\downarrow$) & co-RE-c \cite{Stade,KimPlane}$^*$ & co-RE-c ($\uparrow$)$^*$ & co-RE-c ($\uparrow$)$^*$ \\
    \rowcolor{bgred}\cellcolor{white}%
    4 & \cellcolor{bggreen}R ($\downarrow$) & co-RE-c ($\uparrow$)$^*$ & co-RE-c ($\uparrow$)$^*$ & co-RE-c ($\uparrow$)$^*$ \\
    \rowcolor{bgred}\cellcolor{white}%
    5 & \cellcolor{bggreen}R ($\downarrow$) & co-RE-c \cite{Ollinger} ($\uparrow$)$^*$ & co-RE-c ($\uparrow$)$^*$ & co-RE-c ($\uparrow$)$^*$ \\
    \rowcolor{bgred}\cellcolor{white}%
    $n$ & \cellcolor{bggreen}R \cite{GTTwoTiles} & co-RE-c \cite{Berger,Golomb} ($\uparrow$)$^*$ & co-RE-c ($\uparrow$)$^*$ & co-RE-c ($\uparrow$)$^*$ \\
    \bottomrule
  \end{tabularx}
  \caption{Complexity of tiling Euclidean space by translations and rotations,
    with or without reflections, organized as in \cite[Table~2]{Trominoes}.
    Notation follows \cref{tab:related}.
    }
  \label{tab:rotations}
\end{table}

\section{Proof Overview \label{sec:overview}}

We prove co-RE-hardness by reducing from the problem of Wang tiling,
introduced by \cite{Wang} and proved co-RE-hard by \cite{Berger}.
A \defn{Wang tile} is a unit square with a color on each of its four edges.
Given a set $W$ of Wang tiles,
a \defn{Wang tiling} assigns a tile from $W$ to each grid point in $\Z^2$
such that, when placing those tiles by translation,
touching edges have the same color.
Deciding whether a given finite set of Wang tiles admits such a tiling
is co-RE-complete \cite{Berger}.

Thus our main task for $\Z^3$ is to construct a computable map from a
finite Wang tile set $\calT$
to a finite connected set $F_{\calT}\subset\Z^3$ such that
$\calT$ tiles $\Z^2$ if and only if
$F_{\calT}$ tiles $\Z^3$ by translations.
To achieve this, we combine five constructions,
as summarized in \cref{fig:reduction}:

\begin{figure}[bp]
  \centering
  \begin{tikzpicture}[
    font=\small, >=Latex,
    stage/.style={draw=black!55, rounded corners=2pt, fill=black!3,
      text width=4.0cm, minimum height=1.4cm, align=center},
    arrowlabel/.style={font=\small\bfseries, fill=white, inner sep=2pt},
    reductionarrow/.style={->, semithick, draw=black!65}]
    \node[stage] (wang) at (0,0) {Wang tiling\\of $\Z^2$};
    \node[stage] (sudoku) at (5.5,0) {Sudoku array};
    \node[stage] (system) at (11,0) {Family of prototiles\\in $\Z^2\times(\Z/m\Z)$\\with a common\\set of translations};
    \node[stage] (stacked) at (0,-2.4) {Single prototile\\in $\Z^2\times(\Z/m\Z)$\\with a larger $m$};
    \node[stage] (lattice) at (5.5,-2.4) {Single prototile\\in $\Z^3$};
    \node[stage] (connected) at (11,-2.4) {Single connected prototile\\in $\Z^3$};
    \draw[reductionarrow] (wang) -- node[arrowlabel,above] {1} (sudoku);
    \draw[reductionarrow] (sudoku) -- node[arrowlabel,above] {2} (system);
    \draw[reductionarrow,rounded corners=4pt]
      (system.east) -- (13.5,0) -- (13.5,-1.2)
      -- node[arrowlabel] {3} (-2.5,-1.2)
      -- (-2.5,-2.4) -- (stacked.west);
    \draw[reductionarrow] (stacked) -- node[arrowlabel,above] {4} (lattice);
    \draw[reductionarrow] (lattice) -- node[arrowlabel,above] {5} (connected);
  \end{tikzpicture}
  \caption{The five reductions in our construction for tiling $\Z^3$ by translations.
    Arrow numbers match the steps in the text; every reduction preserves
    existence of a solution in both directions.}
  \label{fig:reduction}
\end{figure}

\begin{enumerate}
\item Greenfeld and Tao \cite[Section~5]{GTUndecidable} reduce Wang tiling
   to finding a \defn{Sudoku array} --- a grid with infinitely many rows
   and a fixed number of columns (independent of the Wang tile set)
   such that every integer-slope line reads as one of a specified finite
   set of words.
   Specifically, each grid symbol specifies a Wang tile together with
   two integers; the Sudoku rule organizes these integers using powers of
   two fixed distinct primes $p,q$, whose exponents encode the two coordinates
   of the simulated plane.
   The line rules compare neighboring positions in this simulated plane
   to enforce matching edge colors.
   We additionally require that neither integer component is constant
   down any column.
\item We adapt OpenAI's cyclic tiling encoding \cite[Sections~3--5]{OpenAI2026}
   to the Sudoku array problem from the previous step.
   (OpenAI originally applied this encoding to a different construction
   of Greenfeld and Tao's \cite[Section~4]{GTUndecidable},
   which uses a one-prime Sudoku rule.)
   This adaptation is our main new work: it enforces the Wang tile constraints
   and both nonconstant-column conditions simultaneously.
   The set of translations encodes the Sudoku array, and the prototiles
   test the rules: forbidden combinations of symbols cause overlaps or gaps.
   The result is a finite family of prototiles (depending on the Wang tile set)
   that each tile a target space using the same infinite set of translations,
   if and only if the required Sudoku array exists.
   The target space being tiled is $\Z^2 \times (\Z/m\Z)$,
   a 3D integer grid whose $z$ coordinate wraps around modulo~$m$
   (where $m$ depends on the Wang tile set).
\item We apply OpenAI's stacking construction \cite[Section~6]{OpenAI2026}
   (based on an earlier construction of Greenfeld and Tao
   \cite[Theorem~3.1 and Lemma~3.2]{GTPeriodic})
   to combine the prototiles from the previous step
   into a single prototile that tiles a (larger) target space
   if and only if such a common set of translations exists.
   This construction enlarges $m$ in the target space $\Z^2 \times (\Z/m\Z)$.
\item We apply the reduction of Meyerovitch, Sanadhya, and Solomon
   \cite[Theorem~1.1]{MSS}, which converts tiling problems in a quotient
   of $\Z^3$ into tiling problems in $\Z^3$,
   while preserving the number of prototiles.%
   \footnote{%
     The proof of \cite[Lemma~2.1]{MSS} has a gap in its local forcing argument:
     covering a neighboring box center without overlap does not force
     the claimed placement. Our Lean formalization gives a counterexample
     to this local claim and proves that their reduction is nonetheless
     correct in our setting.}%
   \ 
   Our target space $\Z^2 \times (\Z/m\Z)$ is such a \defn{quotient}:
   it identifies each point $(x,y,z)$ with $(x,y,z+m)$.
   Their construction uses matching bumps and dents
   to force alignment to a scaled grid and
   to force repetition every $m$ grid steps in the $z$ direction.
   Thus we convert the prototile from the previous step
   into a single finite 3D prototile
   that tiles $\Z^3$ if and only if the original prototile
   tiles the quotient space $\Z^2 \times (\Z/m\Z)$.
\item We use a reduction of Kim \cite[Theorem~2.5]{Kim}
   which makes the prototile connected
   while preserving dimension, the number of prototiles, and tileability.
   This construction scales up the prototile, giving space to connect the
   lattice points by thin corridors, with bumps and dents to force alignment.
\end{enumerate}

For $d\geq3$, appending $d-3$ zero coordinates to each lattice point
preserves tilability, by slicing $\Z^d$ into parallel copies of $\Z^3$.
Replacing each lattice point $p$ by the unit cube $p+[0,1]^d$ then gives a polycube
in $\R^d$ with the same translational tilability; the reverse implication rounds
real translations to a generically shifted lattice.
To preserve tilability under arbitrary Euclidean motions,
we replace each unit cube by a constant-size polycube
with distinct matching keys and recesses on its opposite facets,
which force a common orientation and lattice in every Euclidean tiling.
Co-RE membership, also for finitely many prototiles,
follows by encoding tilings as a computable compact system of
closed polynomial constraints and using finite-precision tests
to recognize infeasibility.

For $n$ Wang tiles, the translation-only reduction uses at most
$B(n)=(n+4)^{10^{61}\lg\lg(n+4)}=n^{O(\log\log n)}$ lattice points / unit cubes
in every fixed dimension $d\geq3$.
For $n=1$, we can improve the upper bound to $2^{4\cdot10^{60}}$.
Allowing rotations and/or reflections gives bounds of $64{,}000B(n)$ in 3D
and $2^{10(d+1)^2}B(n)$ in $d$D.

\section{Formal Verification in Lean \label{sec:lean}}

The proof of \cref{thm:undecidability} is long and technical.
To make it checkable, we formalized it in the Lean~4 proof assistant,
building on Mathlib and on the formalization of 
Wang tiling undecidability (\texttt{lean-{}wang} \cite{LeanWang}).
The complete Lean proof is available in \protect\href{https://github.com/edemaine/translation-tiling-paper/tree/main/lean}{the repository}.
The key point is that a reader does not need to read the proof,
or even most of the definitions, to trust the result:
it suffices to check that the handful of definitions and theorem statements
in a single file, \protect\href{https://github.com/edemaine/translation-tiling-paper/blob/main/lean/TranslationTiling/Abstract.lean}{\texttt{Abstract.lean}}, accurately express the claim.
The Lean proof assistant then verifies the proofs of these statements,
and that they rely on only Lean's three standard axioms
(\texttt{propext}, \texttt{Classical.choice}, \texttt{Quot.sound}).

In the rest of this section, we explain the statements from \protect\href{https://github.com/edemaine/translation-tiling-paper/blob/main/lean/TranslationTiling/Abstract.lean}{\texttt{Abstract.lean}}.
(The source file \protect\href{https://github.com/edemaine/translation-tiling-paper/blob/main/lean/TranslationTiling/Abstract.lean}{\texttt{lean/TranslationTiling/Abstract.lean}}
remains the authoritative version.)
It uses Mathlib's definitions of computability, Euclidean space, affine isometries,
and Lebesgue measure
and \texttt{lean-{}wang}'s definitions of co-RE
(\texttt{CoREPred}, \texttt{CoREHard}, and \texttt{CoREComplete}, defined as in the introduction)
and Wang tiling (\texttt{TileSet} and \texttt{TilesPlane}),
together with a short file \protect\href{https://github.com/edemaine/translation-tiling-paper/blob/main/lean/TranslationTiling/AbstractBridge.lean}{\texttt{AbstractBridge.lean}} whose job is to connect
these statements to the definitions used in the rest of the formalization
(but which need not be checked by a human).

\leansnippet{Imports}

\subsection{Prototiles}

Fix a dimension $d\geq3$. A \defn{prototile} is a finite list $P$ of points in $\Z^d$.
It is a list (not a set) so that it can be given as input to an algorithm;
repetitions are immaterial.

\leansnippet{Point,Prototile}

\subsection{Connectivity}

Two points $a,b\in\Z^d$ are \defn{adjacent} if $\|a-b\|_1=1$.
A prototile $P$ is \defn{connected} if any two points in $P$
are joined by a path of adjacent points inside $P$, i.e.,
they are related by the reflexive transitive closure of
the relation ``both points are in $P$ and adjacent''.

\phantomsection\label{sec:lean-connected}
\leansnippet{Adjacent,Connected}

\subsection{Tiling \texorpdfstring{$\Z^d$}{Zd}}

$P$ \defn{tiles} $\Z^d$ if there exists a set $T\subseteq\Z^d$ of translations
such that every $x\in\Z^d$ lies in exactly one translate $t+P$ with $t\in T$,
i.e., there is exactly one $t\in T$ with $x-t\in P$.
(No periodicity or finiteness assumption is placed on $T$.)

\leansnippet{Tiles}

\subsection{Main Theorem for \texorpdfstring{$\Z^d$}{Zd}}

Deciding whether a connected prototile tiles $\Z^d$ is co-RE-complete
(\texttt{LeanWang.CoREComplete}, from \texttt{lean-{}wang}):
the non-tilers are recursively enumerable, and every co-RE problem reduces to
it by a computable many-one reduction.
Consequently (and using only Mathlib's computable predicate)
no algorithm decides it.
In both statements, the proof is a single call to a theorem in
\protect\href{https://github.com/edemaine/translation-tiling-paper/blob/main/lean/TranslationTiling/AbstractBridge.lean}{\texttt{AbstractBridge.lean}}.

\phantomsection\label{sec:lean-z3}
\leansnippet{tiling_coRE_complete,tiling_undecidable}

A computable reduction converts a list of $n$ Wang tiles into an
equivalent nonempty connected prototile $P\subset\Z^d$ using
at most $B(n) = (n+4)^{10^{61}\lg\lg(n+4)}$ unit cubes:
the Wang tiles tile the plane if and only if $P$ tiles $\Z^d$ by translations.
Here \texttt{LeanWang.TileSet} is a list of Wang tiles, and \texttt{LeanWang.TilesPlane}
asserts that they tile the whole plane.
List lengths count the input tiles and output cubes,
and \texttt{Real.logb\ 2} denotes the base-2 logarithm $\lg$.

\leansnippet{tiling_reduction}

\subsection{Tiling \texorpdfstring{$\R^d$}{Rd}}

We represent a \defn{point} \texttt{x} of $\R^d$ using Mathlib's \texttt{EuclideanSpace},
which provides the usual Euclidean norm.
A coordinate is written \texttt{x\ i} for \texttt{i\ :\ Fin\ d},
just like our \texttt{Point} definition for $\Z^d$.
The \defn{solid body} of $P$ is the union of the closed unit cubes $p+[0,1]^d$ for $p\in P$.

\leansnippet{RealPoint,Solid}

A \defn{Euclidean motion} is an affine bijection preserving distances.
Mathlib's \texttt{AffineIsometryEquiv} represents such a motion. In the notation below,
$\simeq$ denotes a bijection, the superscripts mean affine and isometric,
and $[\R]$ specifies real scalars.

\leansnippet{Motion}

We define four sets of motions: all Euclidean motions; translations alone;
translations and rotations; and translations and at most one arbitrary hyperplane reflection.
A motion is a translation exactly when its linear part is the identity,
and is a translation composed with a rotation exactly when its linear part has determinant $1$.
In Lean, \texttt{f.linear} extracts the linear part, \texttt{1} denotes the identity,
and \texttt{.det} computes the determinant.
Mathlib's \texttt{AffineSubspace.perpBisector\ p\ q} is the set of points equidistant
from \texttt{p} and \texttt{q}, a hyperplane when they are distinct.
The function \texttt{EuclideanGeometry.reflection} reflects in this affine subspace.
The product applies the reflection first, then the translation.
When \texttt{p\ =\ q}, the bisector is the whole space and its reflection is the identity,
so translations alone are included too.

\leansnippet{allMotions,translations,translationsAndRotations,translationsAndReflections}

The predicate \texttt{RealTiles\ P\ motions} requires a set of permitted motions whose
placed copies cover Lebesgue-almost every point exactly once.
The notation \texttt{f\ \textquotesingle{}\textquotesingle{}\ S} means the image $\{f(x):x\in S\}$.
(``Almost every'' so that cubes may share boundaries.)

\leansnippet{RealTiles}

\subsection{Main Theorem for \texorpdfstring{$\R^d$}{Rd}}

Deciding whether a connected polycube tiles $\R^d$ is co-RE-hard
(\texttt{LeanWang.CoREHard}, from \texttt{lean-{}wang}) and undecidable
for any set of Euclidean motions containing all translations.

\phantomsection\label{sec:lean-rigid}
\leansnippet{real_tiling_coRE_hard,real_tiling_undecidable}

For each of the four motion sets defined above, tiling $\R^d$
with a single finite connected polycube is co-RE-complete.

\phantomsection\label{sec:lean-r3}
\leansnippet{real_tiling_coRE_complete}

Together, these Lean theorems establish \cref{thm:undecidability}.

The same computable reduction and cube bound from $\Z^d$ apply.
The following are equivalent: the input Wang tiles tile the plane;
the output prototile tiles $\Z^d$ by lattice translations; and
its solid body tiles $\R^d$ by arbitrary real translations.

\leansnippet{real_tiling_reduction}

A computable reduction produces a nonempty connected polycube that tiles
$\Z^d$ by lattice translations if and only if the input Wang tiles tile the plane.
For any set of Euclidean motions containing all translations, this is also
equivalent to its solid body tiling $\R^d$ using those motions.
Its cube bound is $2^{10(d+1)^2}B(n)$.

\leansnippet{rigid_tiling_reduction}

In 3D, the bound improves to $64{,}000B(n)$:

\leansnippet{rigid_tiling_reduction_three}

\section{Our Approach to the Proof \label{sec:approach}}

We made heavy use of AI to find, write, and formalize the proof,
and we do not claim that this paper is a complete human-readable account of it.
Instead, this paper contains only the part intended for humans:
an explanation of the problem (\cref{sec:intro}),
the main intuition behind the proof (\cref{sec:overview}),
and the precise formalization of the statement (\cref{sec:lean}).
The rigor of the proof comes from the Lean formalization rather than from the text.

A much longer LLM-generated technical article, which is not meant for
human consumption, gives the full details of the construction and its proofs.
It and the complete Lean code are available in
\protect\href{https://github.com/edemaine/translation-tiling-paper}{the repository}
\cite{TilingRepo}
(files \protect\href{https://github.com/edemaine/translation-tiling-paper/blob/main/slop.texlish}{\texttt{slop.texlish}}
and \protect\href{https://github.com/edemaine/translation-tiling-paper/tree/main/lean}{\texttt{lean/}}, respectively).
We intend to expand this paper with further human-readable explanations
of the proof.

\section*{Acknowledgments}

We thank the MIT CompGeom Group for helpful discussions
and for providing an inspiring atmosphere.
This paper was typeset with \protect\href{https://texlish.org/}{Texlish} \cite{Texlish}.
We used the following LLM models:
GPT 6.1 Sol, GPT 6 Astra,
Claude Sonnet 5.5, Claude Opus 5.5, and Claude Fable 5.1.

\AddToHook
   {cmd/thebibliography/after}%
   {\setlength\itemsep{0pt}%
    \setlength\parskip{0pt}}%

\bibliographystyle{alpha-key}%
\bibliography{paper}%

\end{document}